\documentclass[twocolumn,showpacs,prl,amsmath,amssymb]{revtex4}
\usepackage{graphicx}
\usepackage{epsfig}
\usepackage{dcolumn}
\usepackage{amsmath}
\usepackage{bm}
\begin{document}
\title{Rayleigh Imaging of Graphene and Graphene Layers}
\author{C. Casiraghi$^1$, A. Hartschuh$^2$, E. Lidorikis$^3$, H. Qian$^2$,\\ H. Harutyunyan$^2$, T.
Gokus$^2$, K. S. Novoselov$^4$, A. C. Ferrari$^1$}
\email{acf26@eng.cam.ac.uk}
\affiliation{$^1$Cambridge University, Engineering Department, Cambridge, UK\\
$^2$ Chemistry and Biochemistry Department and CeNS,
Ludwig-Maximilians- University of Munich, Germany\\ $^3$ Department
of Materials Science and Engineering, University of Ioannina, Greece\\
$^4$ School of Physics and Astronomy, University of
Manchester,Manchester, M13 9PL, UK.}

\begin{abstract} We investigate graphene and graphene layers on
different substrates by monochromatic and white-light confocal
Rayleigh scattering microscopy. The image contrast depends
sensitively on the dielectric properties of the sample as well as
the substrate geometry and can be described quantitatively using the
complex refractive index of bulk graphite. For few layers ($<$6) the
monochromatic contrast increases linearly with thickness: the
samples behave as a superposition of single sheets which act as
independent two dimensional electron gases. Thus, Rayleigh imaging
is a general, simple and quick tool to identify graphene layers,
that is readily combined with Raman scattering, which provides
structural identification.\end{abstract} \maketitle

Graphene is the prototype two dimensional carbon
system~\cite{geimrev}. Its electron transport is described by the
(relativistic-like) Dirac equation and this allows access to the
rich and subtle physics of quantum electrodynamics in a relatively
simple condensed matter experiment ~\cite{ Novoselov2005proc,
Novoselov2005nature, Zhang2005nature, Novoselov2004, Novoselov2006}.
The scalability of graphene devices to true nanometer
dimensions\cite{kimribbon,Lemme,avouris} makes it a promising
candidate for future electronics, because of its ballistic transport
at room temperature combined with chemical and mechanical stability.
Remarkable properties extend to bi-layer and
few-layers~\cite{Novoselov2004, Novoselov2006, Zhang2005apl,
Berger2004, Scott2005}. More fundamentally, the various forms of
graphite, nanotubes, buckyballs can all be viewed as derivatives of
graphene.

Graphene samples can be obtained from micro-mechanical cleavage of
graphite~\cite{Novoselov2005proc}. Alternative procedures include
chemical exfoliation of graphite ~\cite{Viculis03, Viculis05,
Niyogi06, Stankovich06, Stankovich06JMC} or epitaxial growth by
thermal decomposition of SiC~\cite{Berger2004,Forbeaux99,Ohta06,
Rolling06}. The latter has the potential of producing large-area
lithography compatible films, but is substrate limited. It is hoped
that in the near future efficient large area, substrate independent,
growth methods will be developed, as it is now the case for
nanotubes.

Despite the wide use of the micro-mechanical cleavage, the
identification and counting of graphene layers is still a major
hurdle. Monolayers are a great minority amongst accompanying thicker
flakes~\cite{geimrev}. They cannot be seen in an optical microscope
on most substrates. Currently, optically visible graphene layers are
obtained by placing them on the top of oxidized Si substrates with
typically 300 nm SiO${_2}$. Atomic Force Microscopy (AFM) is viable
but has a very low throughput. Moreover, the different interaction
forces between the AFM probe, graphene and the SiO$_{2}$ substrate,
lead to an apparent thickness of 0.5-1 nm even for a single
layer~\cite{Novoselov2005proc, Novoselov2004}, much bigger of what
expected from the interlayer graphite spacing. Thus, in practice, it
is only possible to distinguish between one and two layers by AFM if
graphene films contain folds or wrinkles~\cite{Novoselov2005proc,
Novoselov2004}. High resolution transmission electron microscopy is
the most direct identification tool~\cite{meyer,acf}, however, it is
destructive and very time consuming, being viable only for
fundamental studies~\cite{meyer}.

Optical detection relying on light scattering is especially
attractive because it can be fast, sensitive and not-destructive.
Light interaction with matter can be elastic or inelastic, and this
corresponds to Rayleigh and Raman scattering, respectively. Raman
scattering has recently emerged as a viable, non-destructive
technique for the identification of graphene and its
doping~\cite{acf,Pisana}. However, Raman scattered photons are a
minority compared to those elastically scattered. Here we show that
the elastically scattered photons provide another very efficient and
quick means to identify single and multi-layer samples and a direct
probe of their dielectric constant.
\begin{figure}
\centerline{\includegraphics[width=70mm]{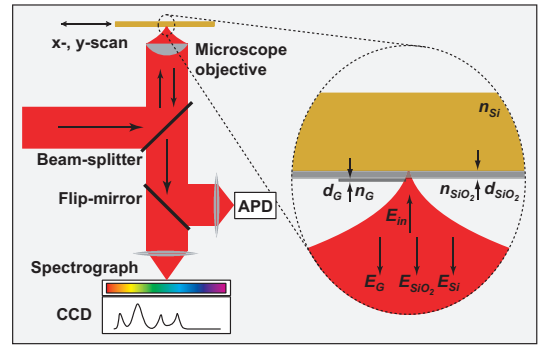}}
\caption{Schematic experimental set-up for combined Rayleigh and
Raman spectroscopy. The inset shows a cross sectional view of the
interaction between the optical field and graphene deposited on Si
covered with SiO${_2}$.} \label{fig1}
\end{figure}

Rayleigh scattering was previously used to monitor size, shape,
concentration and optical properties of nano-particles, carbon
nanotubes  and viruses \cite{lukas, vahid,antonio,tony,tony2}.
Rayleigh scattering experiments can be performed using two different
strategies. In one, the background signal is minimized by making
free-standing samples, as done in the case of carbon nanotubes
\cite{tony,tony2}, or by dark-field configurations \cite{dark}.
Alternatively, the background intensity is utilized as a reference
beam, while the sample signal is detected interferometrically
\cite{lukas, vahid,antonio,kim,calatroni}. Here, we combine the
second approach with the interferometric modulation of the
contributing fields and we show that the presence of a background is
essential to enhance the detection of graphene over a certain
wavelength range.

\begin{figure}
\centerline{\includegraphics[width=40mm]{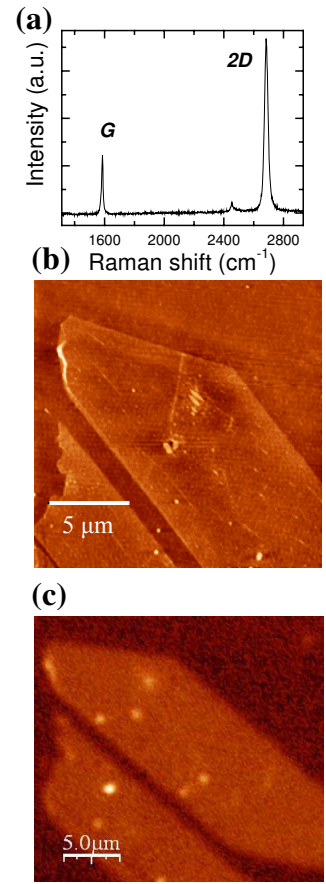}} \caption{(a)
Raman spectrum at 514 nm, showing the features of
graphene~\cite{acf}; (b) AFM image of single layer graphene (c)
Confocal Rayleigh image obtained by raster scanning the sample with
a piezoelectric scan stage.}
 \label{fig3}
\end{figure}

Graphene samples are produced by micro-mechanical cleavage of bulk
graphite and deposited on a Si substrate covered with 300 nm
SiO$_{2}$ (IDB Technologies LTD). The sample thickness is
independently confirmed by a combination of AFM and Raman
spectroscopy. AFM is performed in tapping mode under ambient
conditions. Raman spectra are measured at 514 nm using a Renishaw
micro-Raman 1000 spectrometer. Rayleigh scattering is performed with
an inverted confocal microscope, Fig.\ref{fig1}. Either a He-Ne
laser (633 nm) or a collimated white-light beam are used as
excitation source. Coherent white-light pulses are generated by
pumping a photonic crystal fibre with the output of a Ti:Sa
oscillator operating at 760 nm. The beam is reflected by a beam
splitter and focused by a microscope objective with high numerical
aperture (NA=~0.95). However,the objective lens is not totally
filled, which results in an effective NA$\sim$0.7 thereby increasing
the image contrast as discussed at the end of this paper. The
scattered light from the sample is collected in backscattering
geometry, transmitted by a beam splitter and detected by a
photon-counting avalanche photodiode (APD),~Fig.\ref{fig1}.
Alternatively, the reflected light is filtered using a notch filter
to remove the laser excitation and sent to a spectrometer.
\textit{This allows simultaneous Rayleigh and Raman measurements},
Fig.\ref{fig1},\ref{fig3}a. Confocal Rayleigh images are obtained by
raster scanning the sample with a piezoelectric scan stage. The
acquisition time per pixel varies from few ms in the case of
Rayleigh scattering to few minutes for Raman scattering. This
empirically indicates that Rayleigh measurements are almost 5 orders
of magnitude quicker than Raman measurements. The spatial resolution
is $\sim$ 800 nm.

\begin{figure}
\centerline{\includegraphics[width=60mm]{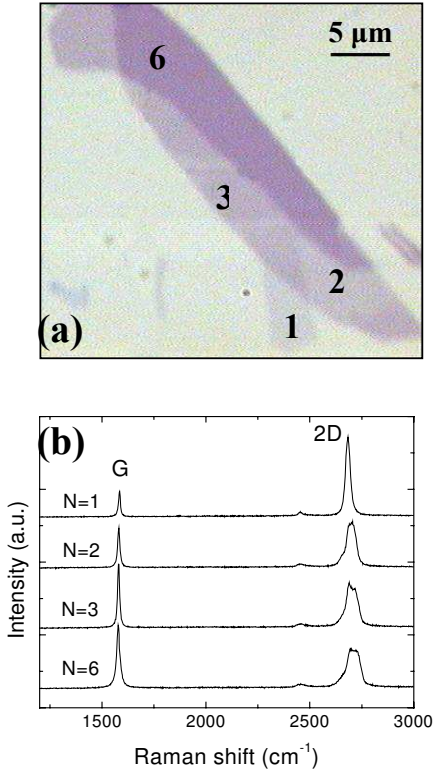}} \caption{(a)
Optical micrograph of multi-layer with 1, 2, 3 and 6 layers; (b)
Raman spectra as a function of number of layers.} \label{fig4}
\end{figure}

\begin{figure}
\centerline{\includegraphics[width=60mm]{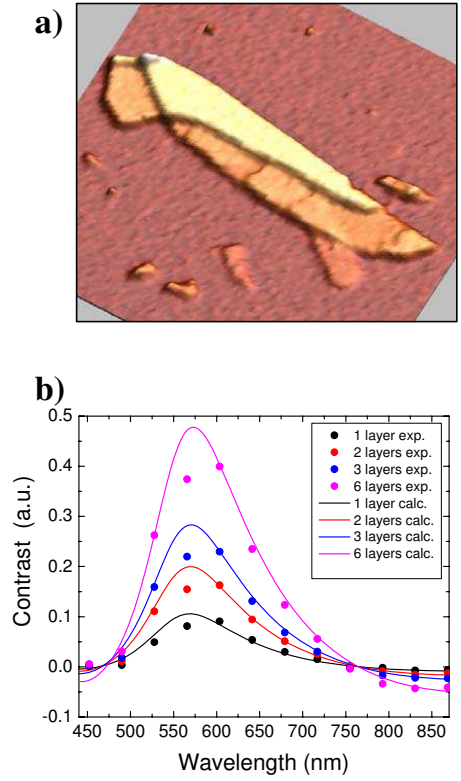}} \caption{(a)
Three-dimensional confocal Rayleigh map for monochromatic 633 nm
excitation. The window size is 49$\mu$m x 49$\mu$m; (b) Experimental
(dots) and theoretical (line) contrast as a function of excitation
wavelength.} \label{fig5}
\end{figure}

Fig.\ref{fig3}(b) shows an AFM image of monolayer graphene. The AFM
cross section gives an apparent height of $\sim$0.6 nm. Raman
spectroscopy confirms that the sample is a single layer
(Fig.\ref{fig3}(a))~\cite{acf}. Fig.\ref{fig3}(b) is the
corresponding confocal Rayleigh image obtained with monochromatic
laser light (633 nm).
Fig.\ref{fig4}(a) shows an optical micrograph of a sample composed
of a varying number of layers. Once the single layer is identified
by Raman scattering, we get the total number of layers from the
measured AFM height, considering the interlayer spacing of
$\sim$~0.33 nm: z [nm]= 0.27 + 0.33 N. This confirms that the sample
is composed of 1, 2, 3 and 6 layers, as for Fig.\ref{fig4}~(a).
These layers have a slightly different color in the optical
microscope (Fig.\ref{fig4}~(a)). It appears that the darker color
corresponds to the thicker sample. Note, however, that the color of
much thicker layers (more than 10 layers) does not follow this trend
and can change from blue, to yellow, to grey. The number of layers
is further confirmed by the evolution of the 514 nm Raman
spectra~\cite{acf},Fig.~\ref{fig4}~(b). Fig.\ref{fig5}(a) shows a
confocal Rayleigh map for 633nm excitation. The signal intensity of
in Fig.\ref{fig5} appears to increase with N.

We now discuss the physical origin of the image contrast ($\delta$).
This is defined as the difference between substrate and sample
intensity, normalized to the substrate intensity. The single layer
contrast at 633 nm is $\sim 0.08$. The contrast is positive, i.e.
the detected intensity from graphene is smaller than that of the
substrate. The Rayleigh images in Fig.\ref{fig3}~(c) and
Fig.\ref{fig5}~(a) are reversed for convenience, in order to compare
them with AFM.

We explain the sign and scaling of the contrast for increasing N in
terms of interference from multiple reflections. The inset in
Fig.\ref{fig1} shows a schematic of the interaction between the
light and graphene on Si+SiO${_2}$. When the light impinges on a
multi-layer, multiple reflections take place \cite{wolf}. Thus, the
detected signal (I) results from the superposition of the reflected
field from the air-graphene ($E_{G}$), graphene-SiO$_2$
($E_{SiO_2}$), and SiO$_2$-Si interfaces ($E_{Si}$). The back-ground
signal ($I_{Bg}$) results from the superposition of the reflected
field from the air-SiO$_2$ interface and the Si substrate.

Before giving a complete quantitative model, it is useful to
consider a simplified picture that captures the basic physics and
illustrates why a single atomic layer can be visualized optically.
The field at the detector is dominated by two contributions: the
reflection by the graphene layer, and the reflection from the Si
after transmission through graphene and after passing through the
SiO${_2}$ layer twice. Thus, the intensity at the detector can be
approximated as:
    \begin{equation}
    I \sim |E_{G}+E_{Si}|^2=|E_{G}|^2 + |E_{Si}|^2 + 2|E_{G}||E_{Si}|\cos\phi
    \label{one}
    \end{equation}
where $\phi$ is the total phase difference. This includes the phase
change due to the optical path length of the oxide, $d_{SiO_2}$, and
that due to the reflection at each boundary,
$\vartheta_{Si}$ and $\vartheta_{G}$:
    \begin{equation}
    \phi= \vartheta_{G}- \left (\vartheta_{Si} + 2\pi~ n_{SiO_2} 2 d
    _{SiO_2}/\lambda_0 \right)
    \label{two}
    \end{equation}
where $n_{SiO_2}$ is the refractive index of the oxide and
$\lambda_0$ is the wavelength of the light in vacuum. Assuming the
field reflected from graphene to be very small, $|E_{G}|^2 \simeq
0$, the image contrast $\delta$ results from interference with the
strong field reflected by the silicon:
\begin{equation}
    \delta=(I_{Si}-I)/I_{Si} \simeq -2\cdot |E_{G} | / |E_{Si}| \cdot
    \cos\phi
    \label{three}
    \end{equation}
The sign of $\delta$ depends on the sign of $\cos\phi$, which is given by
Eq.~\ref{two}.
The reflectance, R, is the ratio between the reflected power to the incident
power \cite{wolf}.
Assuming the Si reflectance as one,~Eq. \ref{three} can be written
as:
    \begin{equation}
    \delta= -2\surd R_{G} ~\cos\phi
    \label{four}
    \end{equation}
where $R_{G}$ is the reflectance of graphene. This is in turn
related to the reflection coefficient $r_{G}$ \cite{wolf}:
    \begin{equation}
    r_G= \surd R_{G} \cdot exp (i \vartheta_{G})
\label{five}
    \end{equation}
Eq.~\ref{four} shows that the main role of the SiO${_2}$ is to act
as a spacer: the contrast is defined by the phase variation of the
light reflected by the Si~\cite{jcp}. \textit{Thus, the contrast for
a given wavelength can be tailored by adjusting the spacer thickness
or its refractive index.}

In order to investigate the wavelength dependence of the image
contrast, we perform Rayleigh spectroscopy with a white-light
source. A grating is used to analyze the detected light.
Fig.\ref{fig5}~(b) shows that for N=1 the contrast is maximum at
$\sim$~570 nm. The contrast at 633 nm is $\sim$~0.08, in agreement
with the monochromatic Rayleigh scattering experiment. The contrast
is zero at 750 nm and it is small and negative for $\lambda$$>$~750
nm. From Eqs.~\ref{two} and ~\ref{four} and assuming
$\vartheta_{Si}=-\pi$, the phase of graphene is $\vartheta_{G}
\simeq -\pi$ as expected for an ultra-thin film~\cite{wolf}. The
contrast decreases in the near IR (for $d_{SiO_2}$=~300 nm) since
the wavelength becomes larger than twice the optical path length
provided by the SiO$_2$-spacer. Fig. \ref{fig5}~(b) shows that while
the contrast increases for increasing N, the phase remains constant.

We now present a more accurate model, with no assumptions, which
describes the light modulation by multiple reflections based on the
recurrent matrix method for reflection and transmission of
multilayered films \cite{hecht}. We calculate the total electric and
magnetic fields in the various layers, applying the boundary
conditions at every interface. The fields at two adjacent boundaries
are described by a characteristic matrix. This depends on the
complex refractive index and the thickness of the film and the angle
of the incident light \cite{hecht}. By computing the characteristic
matrix of every layer and taking into account the numerical aperture
of the objective and the filling factor, it is possible to find the
reflection coefficient for an arbitrary configuration of spacer (2)
and substrate (3) and for any number of graphene layers (G).
Assuming two counter-propagating waves, the standard boundary
conditions for the reflection coefficient of a normally incident
wave is:
\begin{equation}
R=\bigg|\frac{M_{21}}{M_{22}}\bigg|^2 \label{six}
\end{equation}
where
\begin{eqnarray}
M_{12}=\bigg[\cos\phi_{G}\cos\phi_2\bigg(1-\frac{n_{Air}}{n_3}\bigg)-\nonumber\\\sin\phi_{G}\sin\phi_2\bigg(\frac{n_{G}}{n_2}-
\frac{n_{Air}n_2}{n_Gn_3}\bigg)\bigg]\nonumber \\
-i\bigg[\cos\phi_{G}\sin\phi_2\bigg(\frac{n_2}{n_3}-\frac{n_{Air}}{n_2}\bigg)-\nonumber\\\sin\phi_{G}\cos\phi_2\bigg(\frac{n_{G}}{n_3}-
\frac{n_{Air}}{n_{G}}\bigg)\bigg] \label{seven}
\end{eqnarray}
\begin{eqnarray}
M_{22}=\bigg[\cos\phi_G\cos\phi_2\bigg(1+\frac{n_{Air}}{n_3}\bigg)-\nonumber\\\sin\phi_G\sin\phi_2\bigg(\frac{n_G}{n_2}+
\frac{n_{Air}n_2}{n_Gn_3}\bigg)\bigg]\nonumber \\
-i\bigg[\cos\phi_G\sin\phi_2\bigg(\frac{n_2}{n_3}+\frac{n_{Air}}{n_2}\bigg)+\nonumber\\\sin\phi_G\cos\phi_2\bigg(\frac{n_G}{n_3}+
\frac{n_{Air}}{n_G}\bigg)\bigg] \label{eight}
\end{eqnarray}

with $\phi_G=2\pi n_Gd_G/\lambda_0$ and $\phi_2=2\pi
n_2d_2/\lambda_0$. For incidence at an angle $\theta$, with
s-polarization (transverse electric field), the same formula applies
with the substitution $n_i\rightarrow n_i\cos\theta_i$, while for
p-polarization every ratio changes $n_i/n_j\rightarrow
n_i\cos\theta_j/n_j\cos\theta_i$. The phases change in both s and p
polarizations to $\phi_G=2\pi n_Gd_G\cos\theta_G/\lambda_0$ and
$\phi_2=2\pi n_2d_2\cos\theta_2/\lambda_0$. The angle $\theta_i$ for
every layer is obtained from Snell's law:
$\theta_i=\arcsin(\sin\theta_0/n_i)$.  In case any of the layers is
absorbing (as in graphene and Si), we need use an effective index
$n_i'=f(n_i,\theta_0)$ which depends on the incident angle from
vacuum $\theta_0$~\cite{wolf,ciddor}. In this case the corresponding
refraction angle is $\theta_i=\arcsin[\sin\theta_0/Re(n_i')]$.

The matrix method requires as input the complex refractive index of
the sample. The frequency dependent Si and SiO${_2}$ indexes are
taken from Ref. \cite{palik}. For graphene, few layers graphene and
graphite, this is anisotropic, depending on the polarization of the
incident light. For electric field perpendicular to the graphene
c-axis (in-plane) we need $n_{GPerp}$, while for electric field
parallel to the c-axis we need $n_{GParal}$. To get these, we use
the experimental refractive index taken from the electron energy
loss spectroscopy measurements on graphite of
Ref.~\cite{japgraphite}. For s-polarized light (electric field
restricted in the plane) the refractive index to be used is simply
$n_s=n_{GPerp}$. For p-polarization, both in-plane and out-of-plane
field components exists.  Thus we have an angle dependent refractive
index
$n_p^{-2}=n_{GPerp}^{-2}\cos^2\theta_i+n_{GParal}^{-2}\sin^2\theta_i$,
where the refracted angle $\theta_i$ has to be calculated
self-consistently with Snell's law. In order to account for the
numerical aperture in the experiment, we need to integrate the
response of all possible incident angles and polarizations with a
weight distribution accounting for the Gaussian beam profile used in
the experiment
$f(\theta_0)=e^{-2\sin^2\theta_0/\sin^2\theta_m}2\pi\sin\theta_0$,
where $\theta_m=arcsin(NA)$.

Fig.\ref{fig5}(b) shows the calculated contrast for N between 1 and
6 (lines). This is in excellent agreement with the experiments: i)
the contrast scales with number of layers; ii) it is maximum at
$\sim$570 nm; iii) no phase shift is observed in this N range. Thus,
for N between 1 and 6, $\cos \phi (\lambda=570~nm)= -1$. The
contrast of graphene at 570 nm is $\sim$ 0.1. From Eqs.~\ref{four}
and~\ref{five} we get $r_{G}$ ($\lambda$= 570 nm)= 0.05. Thus,
$R_{G}$ ($\lambda$= 570 nm)= 0.003.

It quite remarkable that, without any adjustable parameter,
graphene's response can be successfully modeled using graphite's
dielectric constant. This implies that the optical properties of
graphite do not depend on the thickness, i.e. graphene and graphite
have the same optical constants. The electrons within each graphene
layer form a two dimensional gas, with little perturbation from the
adjacent layers, thus making multi-layer graphene optically
equivalent to a superposition of almost non-interacting graphene
layers. This is intuitive for s-polarization. However, quite notably
this still holds when the out-of-plane direction (p-polarization) is
considered. This is because, compared to the in-plane case,
graphite's response is much smaller, and in addition it gets smeared
out by the NA integration. Thus, the maximum contrast
($\lambda$=~570 nm) of a N-layer is: $\delta(N)=0.1 \cdot N$. Fig.
6(a) shows that this approximation fails for large N. When valid,
the relation between topography and contrast is given by: z[nm]=
0.27 + 3.3$\delta(N)$.
\begin{figure}
\centerline{\includegraphics[width=60mm]{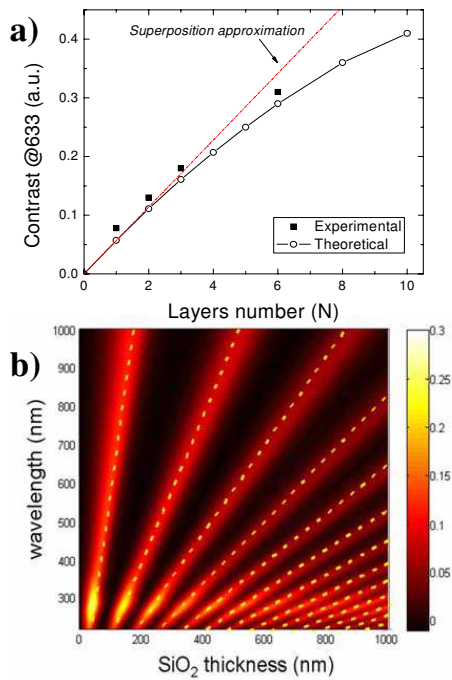}} \caption{(a)
Maximum contrast at 633nm as a function of N; (b) Calculated
contrast of graphene as a function of oxide thickness and excitation
wavelength. Dotted lines trace the quarter-wavelength condition.}
\label{fig6}
\end{figure}

Fig.\ref{fig6}(b) plots the contrast as a function of wavelength and
SiO${_2}$ thickness for a single layer. The maximum contrast occurs
at the minima of the background reflectivity. This is expected
because this is the most sensitive point in terms of phase matching,
and small changes become most visible. Thus, the optimal
configuration requires the SiO${_2}$ to be tuned as an
anti-reflection (AR) coating, i.e. with its optical length a quarter
wavelength. The yellow dotted lines trace the quarter-wave condition
$2n_{Si0_2}d_{Si0_2}/\lambda_0=(m+1/2)$, and indeed they closely
follow the calculated contrast maxima. A second point of interest
are the bright spots around 275 nm. These are due to the absorption
peak at the $\pi\rightarrow\pi*$ transition of graphite
\cite{japgraphite}. For this excitation, the graphene mono-layer not
only becomes much more visible, but the contrast change also
directly reveals the frequency dependence of the graphene's
refractive index. Thus, as for nanotubes~\cite{tony,tony2}, white
light Rayleigh scattering is a direct probe of the dielectric
function.
\begin{figure}
\centerline{\includegraphics[width=60mm]{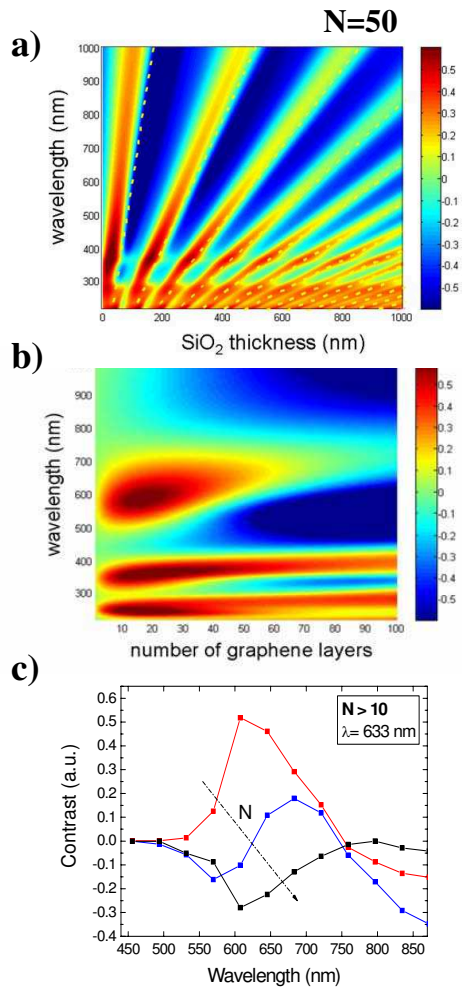}} \caption{(a)
Calculated contrast of 50 layers as a function of oxide thickness
and excitation wavelength; (b)contrast at 633 nm for 300 nm SiO$_2$
as a function of N; (c) Experimental contrast at 633 nm for a thick
sample.} \label{fig7}
\end{figure}

For thicker samples ($N>10$) the phase change due to the optical
path in graphite cannot be neglected. Fig.\ref{fig7}(a) shows the
calculated contrast for a 50 layer sample as a function of SiO$_2$
thickness, while Fig.\ref{fig7}(b) plots the contrast for a fixed
300nm SiO$_2$ thickness, but for a variable number of layers. At 633
nm, as N increases, the response first saturates, then decreases and
red-shifts, finally becoming negative, as found experimentally
(Fig.~\ref{fig7}~(c)). It is also interesting to note that for small
N the variation along the vertical (wavelength) axis is largely
between zero and positive (i.e. reflectivity reduction only), while
for large number of layers, the variation is from positive to
negative (i.e. both reflectivity reduction and enhancement). This
points to two different mechanisms. For small N, the effect of the
graphene layers is just to change the reflectivity of the
air/SiO$_2$ interface, while they offer no significant optical
depth. For large N, on the other hand, the reflectivity of the
air/graphene interface saturates while the effect of the increasing
optical path within the now thick graphite layer becomes
significant. This change is not a monotonic function of N. While
these two effects are different, they both contribute to a shift of
the reflectivity resonance condition, and thus explain the
increasing opaqueness of thicker graphene layers, when measured for
a fixed excitation energy.
\begin{figure}
\centerline{\includegraphics[width=60mm]{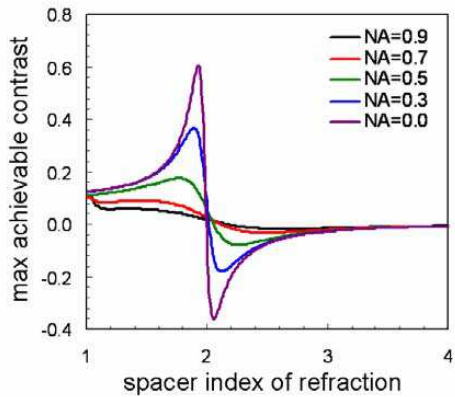}}
\caption{Maximum calculated contrast as a function of spacer
refractive index and objective numerical aperture (NA).}
\label{fig8}
\end{figure}

It is also interesting to consider the contrast as a function of NA.
The calculations show that measurements at a reduced NA would give a
stronger contrast, as one could intuitively expect. However, there
is a nontrivial implication when varying NA, if one tries to
maximize the contrast by using the anti-reflection coating rule for
the spacer. The ideal AR coating over a substrate of index
$n_{subst}$ must have an index $n_{spacer}=\surd{n_{subst}}$ and
quarter wave thickness
$d_{spacer}=(m+1/2)\lambda_0/2\surd{n_{subst}}$. Since $n_{Si}\sim4$
at 600 nm, it is natural to think that a spacer of n=2 (e.g.
Si$_3$N$_4$) would be ideal. To explore this, Fig. \ref{fig8} plots
the contrast for different NAs as a function of $n_{spacer}$ at
600nm and for spacer thickness
$d_{spacer}=300nm(n_{Si0_2}/n_{spacer})$ , which serves to maintain
the AR condition and thus the maximum response.

Contrary to expectations, the contrast maximizes for different
spacer indexes depending on NA. For normal incidence, it is maximum
at 1.93 with a huge contrast of 0.6 for a single layer, Fig.
~\ref{fig8}. It also has a strong variation thereafter, and becomes
negative. As NA further increases, the peak moves to a smaller index
(around 1.5 for NA=0.7), becomes relatively flat, and eventually
goes to $n_{spacer}=1$. Thus, for large NA, it makes little
difference what the spacer index is, as long as the quarter-wave
condition is satisfied. Indeed, for the ideal AR condition the
background reflectivity goes to zero and thus the contrast becomes
large, however this condition strongly depends on the incidence
angle and is thus easily destroyed at large NAs. For all possible
spacer refractive indexes, a reduction in NA results into an
increased contrast, however, the magnitude of this increase varies:
at n=1.5 going from 0.7 to 0.0 NA changes the contrast by a factor
2, while at n=1.9 one can gain a factor of 6, Fig.~\ref{fig8}. For
maximum visibility, a $Si_3N_4$ spacer of thickness 225nm with
NA=0.0 would be ideal. However,if high resolution is needed, as for
nano-ribbons or, in general, to analyze edges and defects, a
compromise between resolution and image contrast is necessary.
\begin{figure}
\centerline{\includegraphics[width=40mm]{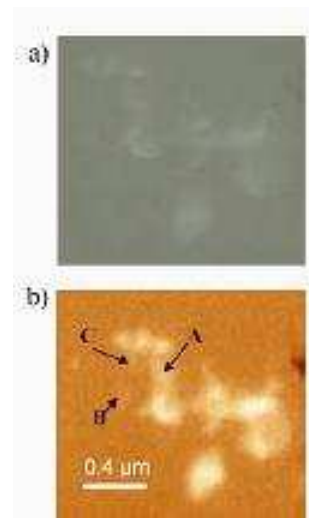}} \caption{(a)
Optical micrograph of flakes on glass. (b) Rayleigh image at 633 nm
excitation. The contrast is much higher compared to
(a).}\label{fig9}
\end{figure}

A second point to note is that for all NAs the contrast converges to
the same value for n=1, i.e. for a suspended graphene layer over the
substrate. Indeed, optically visible suspended layers were recently
reported (see Fig.1 of Ref.~\cite{bunch}). Maximum visibility is
achieved if the quarter-wave condition is satisfied, as indeed in
Ref.~\cite{bunch}, where the 300 nm SiO$_2$ spacer is etched to
create an air gap between graphene and the Si substrate.
Interestingly, in this case any measurement with any NA will yield
the same contrast. The same considerations are relevant for the case
of a thin free-standing spacer (no substrate). By tuning at the low
reflection point (now at half-wavelength) and with an NA=0.0 one
could get fair contrasts. However, as soon as NA increases, the
resonance condition is destroyed and the contrast becomes much
smaller than for the SiO$_2$/Si system.
\begin{figure}
\centerline{\includegraphics[width=70mm]{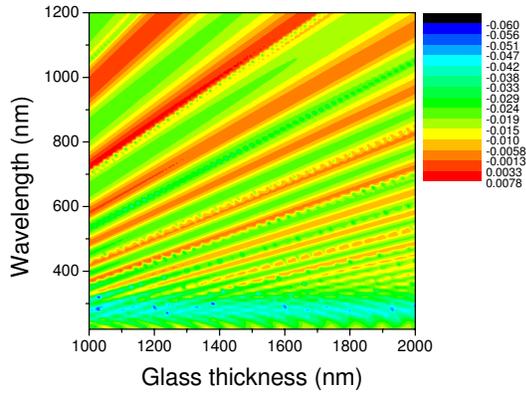}}
\caption{Calculated contrast of graphene on glass at different
wavelengths.}\label{fig10}
\end{figure}

The matrix method can be extended to every film configuration. To
prove this, we measure graphene layers on glass. For N=1, the
calculated contrast at 633 nm is expected to be$\sim$-0.01. Note the
different sign compared with the Si/SiO$_2$ substrate. This is due
to the different optical properties of glass and Si.
Fig.\ref{fig9}(a) shows an optical micrograph of a multi-layer and
Fig.\ref{fig9}(b) the corresponding Rayleigh image at 633 nm. Raman
spectroscopy shows that the sample is composed of layers of
different thickness: A (7-10 layers), B (3-6 layers), C (1-2
layers). Although the contrast is lower compared to Si/SiO$_2$,
Rayleigh spectroscopy allows a better contrast and resolution
compared with the optical microscope, where 10 layers are already
difficult to detect and single layers are practically invisible.
Note that the use of UV light could enhance the contrast to
$\sim$~-0.04 at 300 nm excitation (Fig. \ref{fig10}~(b)).

In conclusion, we used white light illumination combined with
interferometric detection to study the contrast between graphene and
Si/SiO$_2$ substrates. We modeled the light modulation by multiple
reflections, showing that: i) the contrast can be tailored by
adjusting the SiO$_2$ thickness. Without oxide, no modulation is
possible; ii) the light modulation strongly depends on the graphite
thickness. For few layers ($<6$) the samples behave as a
superposition of single sheets. For thicker samples, both amplitude
and phase change with thickness. Thus, Rayleigh spectroscopy
provides a simple and quick way to map graphene layers on a
substrate. It can also be combined with Raman scattering, which is
capable of structural identification.

\paragraph{Acknowledgements.}The authors acknowledge A.K. Geim for useful
discussions. CC acknowledges S. Reich for useful discussions. CC
acknowledges support from the Oppenheimer Fund. ACF from the Royal
Society and Leverhulme Trust. HH from the School of Graduate Studies
G. Galilei (University of Pisa). AH from the German Excellence
Initiative via the Nanosystems Initiative Munich (NIM).

\end{document}